\documentclass[prb, twocolumn,  showpacs, superscriptaddress,  floatfix]{revtex4-1}

\usepackage{latexsym}
\usepackage{amssymb, amsmath}
\usepackage{exscale}
\usepackage{bm, graphics}
\usepackage[final]{graphicx}

\def\bra{\langle}
\def\ket{\rangle}

\begin{document}

\title{Intensity fluctuations in bimodal micropillar lasers enhanced
  by quantum-dot gain competition
}

\author{H.~A.~M.~Leymann}
\affiliation{Institut f{\"u}r Theoretische Physik, Universit{\"a}t
  Magdeburg, Postfach 4120, D-39016 Magdeburg, Germany}
\author{C.~Hopfmann}
\affiliation{Institut f\"{u}r Festk\"{o}rperphysik,
  Technische Universit\"at Berlin, Hardenbergstra{\ss}e 36, D-10623
  Berlin, Germany}
\author{F.~Albert}
\affiliation{Technische Physik and Wilhelm Conrad R\"ontgen Research
  Center for Complex Material Systems, Physikalisches Institut,
  Universit\"at W\"urzburg, Am Hubland, D-97074 W\"urzburg, Germany}
\author{A.~Foerster}
\affiliation{Institut f{\"u}r Theoretische Physik, Universit{\"a}t
  Magdeburg, Postfach 4120, D-39016 Magdeburg, Germany}
\author{M.~Khanbekyan}
\affiliation{Institut f{\"u}r Theoretische Physik, Universit{\"a}t
  Magdeburg, Postfach 4120, D-39016 Magdeburg, Germany}
\author{C.~Schneider}
\affiliation{Technische Physik and Wilhelm Conrad R\"ontgen Research
  Center for Complex Material Systems, Physikalisches Institut,
  Universit\"at W\"urzburg, Am Hubland, D-97074 W\"urzburg, Germany}
\author{S.~H\"ofling}
\affiliation{Technische Physik and Wilhelm Conrad R\"ontgen Research
  Center for Complex Material Systems, Physikalisches Institut,
  Universit\"at W\"urzburg, Am Hubland, D-97074 W\"urzburg, Germany}
\author{A.~Forchel}
\affiliation{Technische Physik and Wilhelm Conrad R\"ontgen Research
  Center for Complex Material Systems, Physikalisches Institut,
  Universit\"at W\"urzburg, Am Hubland, D-97074 W\"urzburg, Germany}
\author{M.~Kamp}
\affiliation{Technische Physik and Wilhelm Conrad R\"ontgen Research
  Center for Complex Material Systems, Physikalisches Institut,
  Universit\"at W\"urzburg, Am Hubland, D-97074 W\"urzburg, Germany}
\author{J.~Wiersig}
\affiliation{Institut f{\"u}r Theoretische Physik, Universit{\"a}t
  Magdeburg, Postfach 4120, D-39016 Magdeburg, Germany}
\author{S.~Reitzenstein}
\affiliation{Institut f\"{u}r Festk\"{o}rperphysik,
  Technische Universit\"at Berlin, Hardenbergstra{\ss}e 36, D-10623
  Berlin, Germany}

\date{\today}

\begin{abstract}
We investigate correlations between orthogonally polarized
cavity modes of a bimodal micropillar laser with a single layer of
self-assembled quantum dots in the active region. While one emission
mode of the microlaser demonstrates a characteristic s-shaped
input-output curve, the output intensity of the second mode saturates
and even decreases with increasing injection current above
threshold. 
Measuring the photon
auto-correlation function $g^{(2)}(\tau)$ of the light emission
confirms the onset of lasing in the first mode with $g^{(2)}(0)$
approaching unity above threshold. In contrast, strong photon bunching
associated with super-thermal values of $g^{(2)}(0)$ 
is detected for the other mode for currents above threshold. 
This behavior is attributed to gain competition of the two modes
induced by the common gain material,
which is confirmed by photon
crosscorrelation measurements revealing a clear anti-correlation
between emission events of the two modes.
The experimental studies are in excellent qualitative
agreement with theoretical studies based on a 
microscopic semiconductor theory, which we   
extend to the case of two modes
interacting with the common gain medium.
Moreover, 
we 
treat the problem by 
an extended birth-death
model
for two interacting modes, which  
reveals, that
the photon probability
distribution of each mode has a
double peak structure,
indicating switching behavior of the modes
for the pump rates
around threshold.    
\end{abstract}

\maketitle

\section{Introduction}
\label{intro}

Quantum dot -- microcavities are a very attractive system to study
quantum 
optical
effects in the solid state~\cite{Rei12}. Apart from
research on fundamental light matter interaction in the weak and
strong coupling regime of cavity quantum
electrodynamics~\cite{Ger98,Bay01,Vah03,Rei04,Yos04}, they offer the
possibility to investigate stimulated emission in a regime approaching
the ultimate limit of a thresholdless laser based on a single 
zero-dimensional gain center~\cite{Nod06}. Studies in this field
include, e.g., technological works on optically and electrically
pumped microlasers aiming at an increase of the $\beta$-factor which
expresses the fraction of spontaneous emission coupled into the lasing
mode~\cite{Wan00,Str06,Rei08}. In high $\beta$-microlasers it becomes
increasingly difficult to identify the transition from spontaneous
emission to stimulated emission at threshold via their input-output
characteristics~\cite{Bjo94}. This issue has triggered comprehensive
experimental and theoretical research activities on the photon
statistics of emission in terms of 
intensity autocorrelation function
in order
to unambiguously identify the onset of stimulated emission at
threshold~\cite{Ric94,Gies:013803,Str06,Ulrich:043906}. Moreover, 
the autocorrelation function
is very beneficial to identify single
quantum dot controlled lasing effects~\cite{Xie07,Rei08a,Nom09} and to
reveal other intriguing effects such as correlations between
individual photon emission events~\cite{Wie09} and chaotic behaviour
of feedback coupled microlasers~\cite{Alb11}.   

The research efforts on microcavity lasers so far have 
focused mostly 
on emission
features based on the interaction between a single laser mode and the
quantum dot gain medium. Going beyond this standard investigations,
micropillar lasers with a bimodal emission spectrum allow one to
address the coupling of two orthogonal optical modes via the common
gain medium which can lead to characteristic oscillations in the
coherence properties~\cite{Ate07} and an enhanced sensitivity on
external perturbations in the presence of optical
self-feedback~\cite{Alb11}. 

In the present work, we perform a detailed
experimental and theoretical analysis of the mode coupling and gain
competition of bimodal, electrically pumped micropillar lasers. 
A convenient measure for the study of
the statistical properties of the electromagnetic field
emission 
is
the set of intensity correlation
functions:
\begin{equation}
  \label{3.11}
  g^{(2)}_{\xi \zeta}(\tau) =
  \frac{
    \langle b^\dagger_\xi (t)b^\dagger_{\zeta}(t+\tau)
    b_{\zeta}(t+\tau) b _\xi(t)
    \rangle
  }
  {\langle b^\dagger_\xi(t) b _\xi(t)  \rangle
    \langle b^\dagger_{\zeta}(t) b _{\zeta}(t)  \rangle },
\end{equation}
where $\xi, \zeta = 1, 2$, with delay time $\tau$ and photon
annihilation operators $b_1$ and $b_2$ of the mode $1$ and the mode
$2$, correspondingly. 
The 
gain competition is reflected in distinct differences in the
input-output characteristic and the autocorrelation function 
$g^{(2)}_{\xi \xi}(\tau)$
of the
two optical modes. 
Moreover, the crosscorrelation function $g^{(2)}_{12}(\tau)$ can
illustrate correlations between emission events from the two modes. 
In order to describe and analysis
these specific features of bimodal microlasers we extend the
microscopic semiconductor model~\cite{Gies:013803} accordingly by taking mode
interactions into account. Similarly, we extend a standard birth-death
approach~\cite{Rice:4318} for the description of bimodal lasers. While
the microscopic semiconductor theory is applied to model the
input-output characteristics, the intensity correlation
functions of the laser and the gain competition
between the two emission modes within a strict mathematical framework,
the extended birth-death approach allows for a more intuitive
understanding of the underlying photon statistics.  

The paper is organized as follows. In section~\ref{sec.1} the experimental results obtained from an electrically pumped, bimodal micropillar laser will be presented. Section~\ref{sec.3} deals with the theoretical description of the experimental data and is divided into two subsections addressing a microscopic semiconductor theory, 
and an extended birth-death approach, respectively.  The paper closes in section~\ref{sec.4} with a comparison of the experimental and theoretical results and a conclusion.

\section{Experiment}
\label{sec.1}

The electrically pumped micropillar lasers are based on planar AlAs/GaAs microcavity structures which includes an active layer consisting of a single layer of In$_{0.3}$Ga$_{0.3}$As. High resolution electron beam lithography, plasma enhanced etching and metal deposition have been applied to fabricate high quality electrically pumped microlasers. For more details on the sample processing we refer to Ref.~\cite{Rei11}. The microlasers have been investigated at low temperature ($20$~K) using a high resolution micro-electroluminescence ($\mu$EL) setup. A linear polarizer in combination with a $\lambda/4$-wave-plate is installed in front of the entrance slit of the monochromator in order to perform  polarization resolved measurements of the laser signal. The photon statistics of the emitted light has been studied by means of the measurement of the  photon autocorrelation function $g^{(2)}_{\xi \xi}(\tau)$, that has been carried out using a fiber coupled Hanbury-Brown and Twiss (HBT) configuration with a temporal resolution $\tau_{\mathrm{irf}}=40$~ps. The HBT configuration is coupled to the output slit of the monochromator which has a focal length of $f = 0.75$~m. The interaction of the orthogonally polarized modes of the microlaser has been investigated by means of photon crosscorrelation measurements. For this purpose, the light emitted by the microlaser is split by a polarization-maintaining 50/50~beamsplitter and coupled into two monochromators ($f = 0.75$~m),
each of which is equipped with a linear polarizer at the input slit and a fiber coupled single photon counting module at the output slit. This configuration allows us to perform polarization resolved crosscorrelation measurements with a spectral resolution of $25~{\mu}$eV.

\begin{figure}[t]
\includegraphics[width=.9\linewidth]{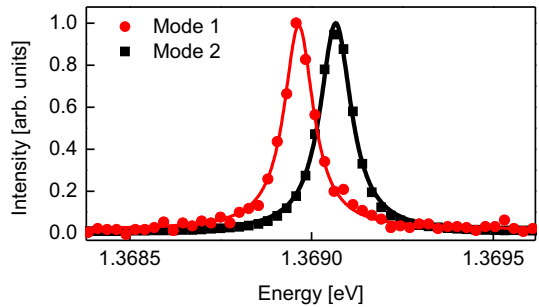}
\caption{\label{fig1} (color online) Polarization resolved ${\mu}EL$ emission spectra of a microlaser with a diameter of $3~{\mu}$m. The electromagnetic field emission features two orthogonally polarized cavity modes, the mode $1$ ($Q = 13900$) and the mode $2$ ($Q = 13100$) with a spectral separation of $103~{\mu}$eV (Injection current: \mbox{$I_{inj}$ $\!=$ $\!5.1~{\mu}$A}).}
\end{figure}

\begin{figure}[t]
\includegraphics[width=.9\linewidth]{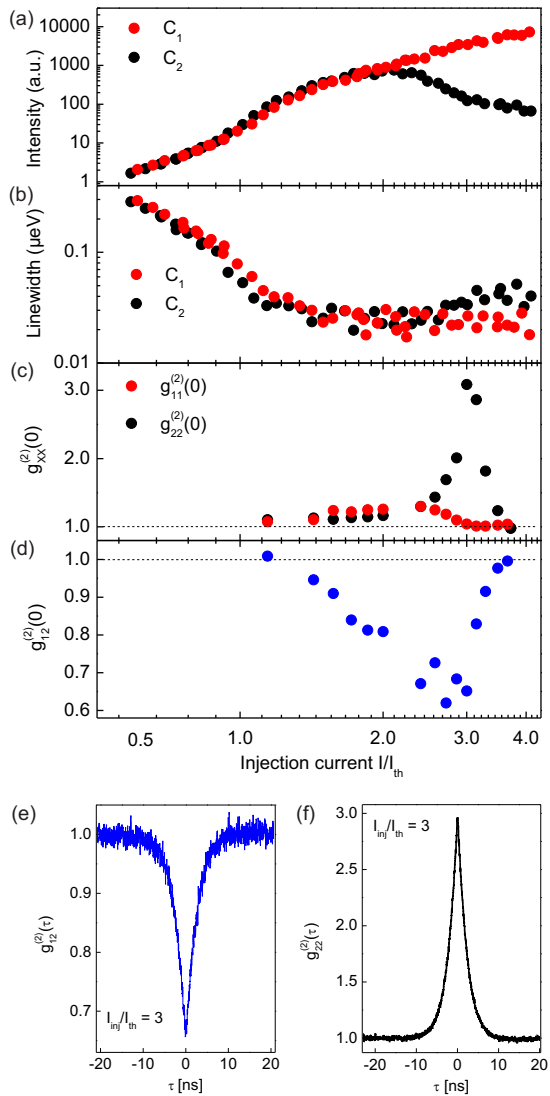}
\caption{\label{fig2} (color online) Experimental characteristics of a bimodal micropillar laser with a diameter of $3~ {\mu}$m. (a) Input-output characteristic,  (b) emission mode linewidth and the photon (c) auto-  and (d) crosscorrelation functions $g^{(2)}_{11}(0)$, $g^{(2)}_{12}(0)$ and $g^{(2)}_{22}(0)$ of emission from modes $1$ and $2$, respectively. Panels (e) and (f) show exemplary crosscorrelation $g^{(2)}_{12}(\tau)$ and autocorrelation $g^{(2)}_{22}(\tau)$ measurements at an injection current of \mbox{$I_{inj}/I_{th}$ = 3}.}
\end{figure}

First, let us focus on the input-output characteristics of the microlaser. Due to slight asymmetry of the cross-section of the pillar and the ring-shaped contact the degeneracy of the fundamental mode in the pillar microcavity is lifted and two distinct linearly polarized modes are supported~\cite{Rei07}.
In this context, the spectral splitting $\Delta_{12}$ and accordingly the overlap between the two modes plays an important role for the studies
of emission of bimodal cavities. Figure~\ref{fig1} shows representative polarization resolved spectra of an electrically pumped bimodal microlaser at threshold 
(injection current,
\mbox{$I_{inj}$ $\!=$ $\!5.1~{\mu}$A}).
The two linearly polarized modes 
are split in energy by $103~{\mu}$eV and have absorption limited $Q$-factors of $Q = 13900$ (mode 1) and $Q = 13100$ (mode 2) at the threshold. The input-output characteristic of the bimodal microlaser is presented in Fig.~\ref{fig2}(a). We observe pronounced differences between the two modes: while mode $1$ shows a standard ``s''-shaped input-output characteristic with a threshold current of about \mbox{$I_{th} = $ $5.1~{\mu}$A}, the intensity of mode
$2$ saturates at \mbox{$I_{inj}/I_{th}$ $\!=$ $\!2$} and even drops down for injection currents exceeding \mbox{$I_{inj}/I_{th}$ $\!=$ $\!2.5$}. This behavior indicates a pronounced competition between the modes $1$ and $2$ which is mediated by the common QD gain material as it will be further elaborated in the following.

Further, to study the lasing features we extract the emission linewidths of the two modes and plot them as a function of the injection current in Fig.~\ref{fig2}(b). The linewidths of the modes $1$ and $2$ have similar magnitude and decrease strongly at threshold which reflects enhanced temporal coherence in the lasing regime. Interestingly, while the linewidth of the mode $1$ stays at a resolution limited value of $25~{\mu}$eV, a slight increase of the linewidth can be observed for the mode $2$ above \mbox{$I_{inj}/I_{th}$ = 3}. This is in agreement with the decreasing emission intensity seen in Fig.~\ref{fig2}(a), which indicates an increasing contribution of spontaneous emission in the mode $2$ at high injection currents.

In order to verify the interpretation of mode coupling in terms of
gain competition, we have performed crosscorrelation measurements
between the modes $1$ and $2$ at different injection currents. An
illustration of such a measurement is presented in Fig.~\ref{fig2}(e)
for \mbox{$I_{inj}/I_{th}$ = 3}. The cross-correlation function
$g^{(2)}_{12}(\tau)$ shows a pronounced dip \mbox{$g^{(2)}_{12,min}$
  $\!=$ $\!0.62$} at \mbox{$\tau$ $\!=$ $\!0$} which indicates an
anti-correlation between emission events from the two laser modes. The
anti-correlated emission occurs at a characteristic timescale of 
\mbox{$\tau_{12}$ $\!=$ $\!3.8$}~ns. Figure~\ref{fig2}(d) reveals
that the crosscorrelation function $g^{(2)}_{12}(0)$ strongly depends
on the injection current.  
It is useful to note, that, as it is seen from Fig.~\ref{fig2} in the
regime of certain injection currents above the threshold 
(\mbox{$2.7$ $\!<$ $\!I_{inj}/I_{th}$ $\!<$ $3.3$}), the intensity of mode $2$ decreases and the statistics of mode $2$ demonstrates strongly super-Poissonian behavior, whereas the anti-correlation between the modes is the strongest.

The interplay between the two emission modes is also accompanied by strong temporal intensity fluctuations which are identified by measuring the photon autocorrelation function of the two competing modes for different injection currents.  The respective dependencies, i.e. $g^{(2)}_{11}(0)$ and $g^{(2)}_{22}(0)$ versus injection current, are plotted in Fig.~\ref{fig2}(c), while Fig.~\ref{fig2}(f) shows the autocorrelation function $g^{(2)}_{22}(\tau)$ for
\mbox{$I_{inj}/I_{th}$ = 3}. The mode $1$ shows the typical maximum of $g^{(2)}_{11}(0)$ around threshold, which indicates the transition from spontaneous emission to stimulated emission, where $g^{(2)}_{11}(0)$ is lower than expected from theory due to the limited temporal resolution of the
HBT~\cite{Ulrich:043906}. In contrast, the autocorrelation function
of the mode $2$, $g^{(2)}_{22}(0)$ increases strongly at the pump
rates well above threshold and reaches a maximum value of $3.08$ at \mbox{$I_{inj}/I_{th}$ = 3}. This value is significantly higher than $g^{(2)}_{22}(0) = 2$, expected for thermal light and, therefore, can
not be explained by 
standard photon statistics.

It is important to note, that similar statistical properties of the emission, i.e. strong super-Poissonian behavior for the weak mode, has been also observed for microlasers in the presence of an external mirror, where a delayed feedback of the emitted signal disturbs laser operation and leads to
strong bunching for the weak mode~\cite{Alb11}.


\section{Theory}
\label{sec.3}

To develop a theoretical framework for the study of the coupled
carrier-photon system in the bimodal cavity we consider two different
theoretical approaches. In the first, a microscopic theory of
light-matter interaction of semiconductor QDs with the cavity field is
given, which allows the derivation of the equations of motion for
quantities of interest. In the second approach, starting with the
master equation, statistics of the photon distribution can be derived
for the case of two-level carriers. 

\subsection{Microscopic Semiconductor Theory}
\label{sec.3.1}

To study the interaction of QDs with the electromagnetic field inside
an optical bimodal microcavities we have extended the Microscopic
Semiconductor Theory~\cite{Gies:013803} to the case of two modes and
photon crosscorrelation functions. 

\subsubsection{Physical Model}
\label{sec.3.1.1}

The Microscopic Semiconductor Theory allows for inclusion of many-body
effects of the carriers and can be used to calculate correlations
required to determine the statistics of the emission of microcavities
with active QDs (for a review see, e.\,g., Ref.~\cite{Gies:1}). The
calculations are based on the cluster expansion truncation scheme of
the equations of motion for operator expectation
values~\cite{Fricke:1996479}.

In what follows we assume that only two confined QD shells for both
electrons and holes are relevant: whereas the resonant interaction
with the electromagnetic field of the bimodal cavity is due to the
coupling with the $s$-shell transition, the carrier generation due to
electrical pumping is to take place in the $p$-shell. 
The assumption suits well also for an experimental situation, where
the electrical pumping is to take place via injection of electrons and
holes into the wetting layer and subsequent fast relaxation to the
discrete electronic states of the QDs. Further, carrier-carrier and
carrier-phonon scattering contributions to the dynamics are evaluated
using a relaxation time approximation, where the relaxation towards
quasi-equilibrium is given in terms of a relaxation
rate~\cite{Nielsen:235314}.

To be more specific, let us consider a bimodal microcavity with the
QDs as gain medium with the driving performed by the 
recombination of carriers in the valence and conduction bands. 
%
The Hamiltonian that governs the temporal evolution of the overall
system can be given in the form   
\begin{equation}
  \label{3.1}
  H = H^0_\mathrm{carr}+H_\mathrm{Coul}+H_\mathrm{ph}+H_\mathrm{D},
\end{equation}
where $H^0_\mathrm{carr}$ is the single-particle contributions for
conduction and valence band carriers  with the energies $\varepsilon_j^{c,v}$,
\begin{equation}
  \label{3.2}
  H^0_\mathrm{carr}=\sum_j\varepsilon_j^{c}c_j^\dagger c_j
  +\sum_j\varepsilon_j^{v}v_j^\dagger v_j,
\end{equation}
and the two-particle Coulomb interaction is given by~\cite{Baer:411}
\begin{multline}
  \label{3.3}
  H_\mathrm{Coul} = \frac{1}{2}\sum_{k ' j j ' k}
  (V^{cc}_{k ' j j ' k}
  c_{k '}^\dagger c_{j}^\dagger c_{j '} c_{k}
  +V^{vv}_{k ' j j ' k}v_{k '}^\dagger
  v_{j}^\dagger v_{j '} v_{k}
)
\\
+
\sum_{k ' j j ' k}
  V^{cv}_{k ' j j ' k}c_{k '}^\dagger v_{j}^\dagger v_{j '} c_{k}.
\end{multline}
In the above,
$c_j$ ($c_j^\dagger$) and $v_j$ ($v_j^\dagger$)
are fermionic operators that annihilate (create) a
conduction-band carrier in the state $| j \ket_c$ and a
valence-band carrier in the state $| j \ket_v$, respectively.
Further, the Hamiltonian of the electromagnetic field modes inside the
cavity reads
\begin{equation}
  \label{3.4}
H_\mathrm{ph} = \sum_\xi \hbar \omega_\xi b_\xi^\dagger b_\xi,
\end{equation}
where $b_\xi$ ($b_\xi^\dagger$) is the bosonic annihilation (creation)
operator of the $\xi$th mode of the cavity.

The energy of interaction of the QDs with the electromagnetic field
inside the cavity in dipole approximation can be given by:
\begin{equation}
  \label{3.5}
  H_\mathrm{D} = -i \sum_{\xi ,  j}
  (g_{\xi j}c_j^\dagger v_j b_\xi + g_{\xi j}v_j^\dagger c_j
  b_\xi)
  +\mathrm{H.c.},
\end{equation}
where
the approximation of equal
wave-function envelopes for conduction- and valence-band states is
used. Moreover, for simplicity the coupling strength $g_{\xi j}$ is
assumed to be real. 

The Hamiltonian given by Eq.~(\ref{3.1}) together with
Eqs.~(\ref{3.2}--\ref{3.5}) determines the dynamical evolution of the
carrier and field operators
and, in particular, the time evolution for operator expectation values.
The equations of motion for quantities of interest, as
for example the average photon number in the cavity modes and the
average electron population in the conduction and valence bands,
have source terms that contain operator expectation values of higher
order. In this way, the approach bears an infinite hierarchy of equations of
motion for various expectation values for photon and carrier operators. To
perform a consistent truncation of the equations the cluster expansion
scheme is applied (for details, see Ref.~\cite{Gies:013803} and
references therein).
Namely, starting from the expectation values of the first order of photon
operators, the equations of motion for
operator expectation values
are replaced by equations of motion for correlation functions.
For example, instead of the equations of motion for expectation values
of amplitudes of
the cavity mode operators $\langle b^\dagger_{\xi} b_{\zeta}\rangle$,
the equations of motion for corresponding amplitude correlation functions
\mbox{$\delta \!\langle b^\dagger_{\xi} b_{\zeta}\rangle$ $\!=\langle
  b^\dagger_{\xi} b_{\zeta}\rangle$ $\!-\langle b^\dagger_{\xi}\rangle
  \langle b_{\zeta}\rangle$}
are used.
Then,
to achieve a consistent classification and inclusion of correlations
up to a certain order
the truncation of the equations for correlation functions rather than
for expectation values is performed.

In particular, in the case of a system without coherent external
excitation
\mbox{$\langle b^\dagger_{\xi} \rangle $ $\!=\langle b_{\xi} \rangle$
  $\!=0$}
and
\mbox{$\langle c^\dagger_{j} v_{j'} \rangle$ $\!=0$}
hold.
Therefore,
applying rotating-wave approximation here and thereafter,
Heisenberg equations of motion for amplitude correlation
functions of the mode operators can be given by
\begin{multline}
  \label{3.6}
  \frac{d}{dt} \delta \! \langle b^\dagger_{\xi} b_{\zeta}\rangle=
  - (\kappa_\xi+ \kappa_\zeta)
  \delta \! \langle b^\dagger_\xi b_\zeta \rangle
\\
  +\sum_{j,q}
  \left(
  g_{\xi j}
    \delta\! \langle c^\dagger_j v_j b_\xi \rangle
    +g_{\xi j}
    \delta \! \langle v^\dagger_j c_j b^\dagger_\zeta \rangle
  \right)
,
\end{multline}
where $\kappa_\xi$ is the loss rate of the $\xi$th cavity mode and
\mbox{$q$ $\!=1$ $\!\dotsc N$}, with $N$ being the total number of QDs.
Note, that
both cavity-mode
amplitude correlation
functions
$\delta \! \langle b^\dagger_\xi b_\zeta\rangle$
and the coupled
photon-assisted polarization amplitude correlations
$\delta \! \langle v^\dagger_j c_j b^\dagger_\xi \rangle $ and
$\delta\! \langle c^\dagger_j v_j b_\zeta \rangle $
are classified as
doublet terms in the cluster expansion scheme, i.e., they
correspond to
an excitation of two electrons (four carrier operators).
The equation
of motion for the photon-assisted polarization amplitude correlation
read [see also Eq.~(\ref{app.7})  in Appendix~\ref{app:1}]:
\begin{multline}
  \label{3.7}
  \frac{d}{dt} \delta \!\langle v^\dagger_{j} c_{j}
  b^\dagger_\xi\rangle=
  -i(\Delta_{\xi j}-i \kappa_\xi-i\Gamma )
  \delta \! \langle v^\dagger_j c_j b^\dagger_\xi \rangle
\\[.5ex]
  +g_{\xi j }  \delta \!\langle c^\dagger_j c_j \rangle
  (
  1 - \delta \! \langle v^\dagger_j v_j \rangle
   )
  +
  \sum_{\xi '}
  \left[
  g_{\xi ' j } \delta \! \langle b^\dagger_{\xi '} b_{\xi}\rangle
  (
  \delta \!\langle c^\dagger_j c_j \rangle
    - \delta \! \langle v^\dagger_j v_j \rangle
    )
  \right.
 \\[.5ex]
  \left.
  +g_{\xi ' j}  \delta \! \langle c^\dagger_j c_j b^\dagger_{\xi ' }
  b_{\xi} \rangle
  -g_{\xi 'j}  \delta \! \langle v^\dagger_j v_j b^\dagger_{\xi '} b_{\xi} \rangle
  \right]
  ,
\end{multline}
where $\Delta_{\xi j}$ $= i(\varepsilon_j^v -\varepsilon_j^c)-\hbar
\omega_\xi  $ is the detuning
of the
$\xi$th cavity-mode from the QD
transition and
$\Gamma$ is a phenomenological dephasing parameter describing
spectral line broadening.
In the case of a bimodal cavity
only the
cavity modes with indices $\xi=1,2$ are resonantly coupled to the
QDs. Whereas the modes with $\xi\neq 1,2$ are not within the gain spectrum of
the QD ensemble or have low $Q$-value. Since the population of the
non-lasing modes
$ \langle b^\dagger_\xi b_\xi \rangle $
and  the cross-correlation
functions $ \langle b^\dagger_\xi b_1 \rangle $ and
$\langle b^\dagger_\xi b_2 \rangle $
with $\xi\neq 1,2$ remain negligibly small,
the third
terms on the right-hand side of Eq.~(\ref{3.7}) can be effectively set
equal to zero. Thus, Eq.~(\ref{3.7}) for $\xi \neq 1,2$ can be solved in
the adiabatic limit yielding a time constant $\tau _{nl}$ that
describes the spontaneous emission into non-lasing modes according to the
Weisskopf-Wigner theory. The spontaneous emission of QDs into
non-lasing modes leading to a loss of excitation is described by
$\beta$-factor defined as the ratio of the spontaneous emission rate into
the lasing modes $1/\tau _{l}$ and the total spontaneous
emission rate enhanced by the Purcell effect $1/\tau _{sp}$:
\begin{equation}
  \label{3.8}
  \beta =
  \dfrac{\tau _{l}^{-1}} {\tau _{sp}^{-1}}
=
  \dfrac{\tau _{l}^{-1}} {\tau _{l}^{-1}+ \tau _{nl}^{-1}}.
\end{equation}

The dynamics of carrier population of the electrons in the $s$-shell
is given by
\begin{multline}
  \label{3.9}
  \frac{d}{dt} \delta \!\langle c^\dagger_s c_s\rangle
  =-\left( \sum_\xi g_{\xi q}  \delta \! \langle c^\dagger_s v_s b_\xi \rangle
    +\mathrm{H.c.}\right )
\\[.5ex]
  +
  \delta \! \langle c^\dagger_p  c_p \rangle
  (1- \delta \! \langle c^\dagger_s  c_s \rangle)
  \tau_c^{-1}
\!-
  \delta \! \langle c^\dagger_s c_s \rangle
   (1- \delta \! \langle v^\dagger_sv_s \rangle)
   \tau_{nl} ^{-1} .
\end{multline}
Here, the first term on the right-hand side originates from the interaction
with the cavity-modes, the second term describes the relaxation of
carriers from the $p$- to the $s$-shell
with a relaxation timescale $\tau _c$
and the term represents the loss of
excitation into the non-lasing modes.

Further, we assume, that the $p$-shell carriers are generated at a
constant pump rate $p$.
Then, similar to Eq.~(\ref{3.9}), the equation of motion for the
carrier population of the electrons in the $p$-shell reads:
\begin{multline}
  \label{3.10}
  \frac{d}{dt} \delta \!\langle c^\dagger_p c_p\rangle
  =p( \delta \! \langle v^\dagger_p  v_p\rangle- \delta \! \langle c^\dagger_p  c_p\rangle )
\\[.5ex]
  -
  \delta \! \langle c^\dagger_p  c_p \rangle
  (1- \delta \! \langle c^\dagger_s  c_s \rangle)
  \tau_c^{-1}
  \!-
  \delta \! \langle c^\dagger_p c_p \rangle
  (1- \delta \! \langle v^\dagger_pv_p \rangle)
  \tau_{sp} ^{-1}
,
\end{multline}
where the last term on the right-hand side describes spontaneous
recombination of $p$-shell carriers. The corresponding equations for
valence band carriers are relegated into Appendix~\ref{app:1}.

The form of the expression for the intensity correlation
functions suggests [see~Eq.~(\ref{3.11})] that to exploit the statistical 
properties of the light emission using intensity correlations, a
consistent treatment within the cluster expansion up to the quadruplet
order is required. In particular, the equations of motion for
cavity-mode intensity correlations read:
\begin{multline}
  \label{3.12}
  \frac{d}{dt} \delta \!\langle b^\dagger_{\xi} b^\dagger_{\xi'}
  b_{\zeta} b_{\zeta '}\rangle
  =
-(\kappa_{\xi}+\kappa_{\xi'}+\kappa_{\zeta}+\kappa_{\zeta '})
  \delta \!
  \langle b^\dagger_\xi b^\dagger_{\xi'} b_{\zeta} b_{\zeta '} \rangle
  \\
  +\sum_{j }
  \left(
  g_{\xi j}  \delta \! \langle c^\dagger_j v_j b^\dagger_{\xi '} b_{\zeta} b_{\zeta '} \rangle
  +g_{\xi ' j}  \delta \! \langle c^\dagger_j v_j b^\dagger_\xi b_{\zeta} b_{\zeta '} \rangle
  \right.
  \\
  \left.
  +g_{\zeta j}  \delta \! \langle v^\dagger_j c_j b^\dagger_\xi b^\dagger_{\xi '} b_{\zeta '} \rangle
  +g_{\zeta ' j}  \delta \! \langle v^\dagger_j c_j b^\dagger_\xi b^\dagger_{\xi '} b_{\zeta} \rangle
  \right).
\end{multline}
The equations of motion for further correlation functions of the
quadruplet order, which include correlation between the
photon-assisted polarization and the photon number, can be found in
Appendix~\ref{app:1} [see Eqs.~(\ref{app.11})--(\ref{app.14})].

\subsubsection{Results}
\label{sec.3.1.2}

As described above, the quadruplet order of the cluster expansion
leads to a system of coupled equations [see
Eqs.~(\ref{3.6})--(\ref{3.7}), (\ref{3.9})--(\ref{3.10}), (\ref{3.12})
together with Eqs.~(\ref{app.7})--(\ref{app.14})]. The system of
differential equations describe the dynamics of 
various correlations between carriers and cavity modes. In particular,
the method makes it possible to obtain both amplitude and intensity
correlation functions of the cavity emission modes including the
effects of the carrier-photon correlations and the many-body Coulomb
interaction. 

In the ensuing section the numerical analysis of the time evolution of
the emission correlation functions is presented. 
To relate our theory to the experimental results 
we estimate the number of QDs with effective gain contribution by
starting with the initial density of present QDs and
excluding the ones with negligible spectral and spatial overlap.
Thus,
it is assumed that the 
cavity mode field is coupled to 
$N$ identical QDs.
Further, we consider continuous carrier generation in the
$p$-shell at a constant rate $p$ as an excitation process. 

\begin{figure}[t]
\includegraphics[width=.9\linewidth]{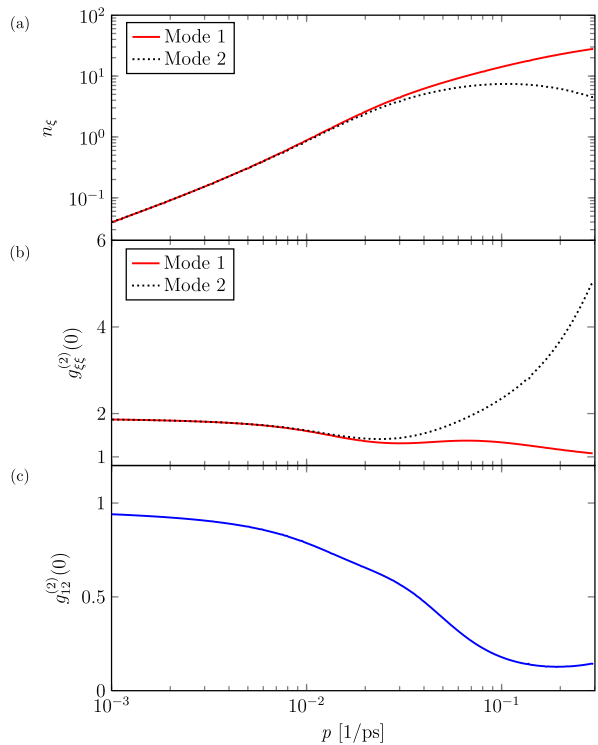}
\caption{\label{fig3} (color online)
  Laser characteristics calculated
  with the semiconductor model.
  (a) Intensity correlation functions for the modes $1$ and $2$ as a
  function of the pump power in a log-log plot.
  (b) Autocorrelation functions of the two modes.
  (c) Crosscorrelation function between the modes $1$ and $2$.
   }
\end{figure}

To obtain a valid comparison with the experimental results we simulate
the coupled system using numerical integration routines with a
realistic set of parameters $\beta = 0.1$, \mbox{$\kappa_1$
  $\!=\kappa_2$ $\!=0.03 \,[1/\mathrm{ps}]$}, $\Gamma =2.06 \,[1/\mathrm{ps}]$, $\tau_{sp} =
50 \,[\mathrm{ps}]$, $\tau_c = 1\,[\mathrm{ps}]$ and
$\tau_{v}=0.5\,[\mathrm{ps}]$.
The number of carriers within the frequency region of interest is
estimated from the total density of QDs to be $N=40$.
For the assumed $\beta = 0.1 $ the carrier
recombination is determined by the stimulated emission into the lasing
modes $1$ and $2$ with a characteristic time scale \mbox{$\tau_{l}$
  $\!=\tau_{sp}/\beta$} and into the non-lasing modes with a
characteristic time scale that can be found from Eq.~(\ref{3.8}) for
the given set of parameters. Further, we assume that the cavity mode
$1$ is in exact resonance with the QD transition 
($\Delta _{1s}=0$)
and the mode $2$ is
detuned with
\mbox{$\Delta _{12}$ $\!\equiv\omega_1-\omega_2$ $\!=\Delta _{2s}$ $\!=0.2 \,[1/\mathrm{ps}]$}. 
In Fig.~\ref{fig3} we present the simulation results for intensity
functions for the modes 
$n_\xi = \bra b^\dagger _\xi b_\xi \ket$, $\xi = 1,2$,
autocorrelation functions and crosscorrelation as a function of the
pump power. Figure~\ref{fig3}(a) reveals, that whereas the mode $1$
shows a drastic increase of emission intensity, the intensity of the
emission mode $2$ reaches a maximum and then slowly decreases with
increasing pump power in good agreement with the experimental
data depicted in Fig.~\ref{fig2}(a). The calculations further show,
that, 
in agreement with the 
experimental data in Fig.~\ref{fig2}(c),
the dependencies of the autocorrelation functions for the cavity
modes $1$ and $2$  on the pump power exhibit dramatically
different behavior. As shown in Fig.~\ref{fig3}(b) for low values of
pump power, the autocorrelation function is equal to $2$
characteristic for the 
statistics of thermal light. For higher rates of the pump power, the
autocorrelation function of the mode $1$ drops close to the value $1$
indicating the emission of 
coherent laser light. In contrast, the autocorrelation function of
mode $2$ slightly decreases at first with increasing pump powers, but
for larger values of the pump power, it increases and reaches values well
above $2$, which is in agreement with the behavior of the
autocorrelation function detected in the experiment \{see
Fig.~\ref{fig2}(c); recall the limited temporal resolution of the
HBT~\cite{Ulrich:043906}\}. 
The gain competition behavior between the modes can be
approved by plotting the crosscorrelation function [see
Fig.~\ref{fig3}(c)], that decreases to the values smaller than unity
at the power pump values for which the lasing behavior of the mode $1$
is observed [also, compare to Fig.~\ref{fig2}(d)]. 
Note, that the discrepancy of the experimental and theoretical results
for the autocorrelation function of mode $2$ [Figs.~\ref{fig2}(c) and
\ref{fig3}(b), correspondingly] and the crosscorrelation function
[Figs.~\ref{fig2}(d) and \ref{fig3}(c), correspondingly] at the higher
pump powers is due to the crosstalk between the modes, which cannot be
completely avoided in the measurements. 

The numerical simulation of the cluster expansion truncation scheme of
the quadruplet order can be approved by plotting the emission mode
autocorrelation 
functions for higher order of truncation (not shown), which
demonstrates qualitatively the same behavior of the functions
independent of the order of truncation. 
It is important to note, that since
the framework of the microscopic semiconductor theory presented in
this section is based on the truncation of the hierarchy of equations for
correlation function, the numerical results are valid
in the regime when higher order correlations remain small. 
As it can seen from the numerical evaluation of the truncated
equations, 
this is not the case for pump power rates exceeding 
$2 \times 10^{-2}\,[1/\mathrm{ps}]$, where the correlation functions
strongly increase. 
To get a deeper understanding of the statistical properties of 
the emission in the next section we will 
use a different approach
to gain
insight into the full photon statistics.

\subsection{Extended Birth-Death Approach}
\label{sec.3.2}

In the following subsection we present an alternative approach to the
study of the light-matter interaction of QDs with a bimodal cavity,
that involves numerically solving a complete master equation and thus
deriving the time evolution of the system. In contrast to the
Microscopic Semiconductor Model discussed in detail in
Sec.~\ref{sec.3.1}, the approach allows to calculate not only photon
auto- and crosscorrelation-functions but also full photon
statistics. We follow the Rice and Carmichael approach (see
Ref.~\cite{Rice:4318}) and extend it to the case of a bimodal cavity
with two resonant modes 
containing $n_1$ and $n_2$ photons, respectively. The method
simplifies the model for the gain medium and takes into account only
fully inverted two-level systems. Note, that no semiconductor effects
or complex level structure are reflected. The state of the gain medium
is fully described by the number of excited carriers $N$. A detailed
discussion of the master equation approach, the semiclassical rate
equations and its connection to the semiconductor theory 
for the case of a single-mode microcavity can be found in
Refs.~\cite{Gies:013803,Gies}
and references therein. The master equation describes the time
evolution of the diagonal elements 
\begin{equation}
  \label{3.2.1}
 \rho^{n_1,n_2}_N=\langle n_1,n_2,N|\rho|n_1,n_2,N\rangle
\end{equation}
of the density matrix $\rho$. These elements can be interpreted as the
probability of finding a state with $n_1$, $n_2$ photons in the modes
$1$ and $2$, respectively, and $N$ Atoms in the excited state. 

\subsubsection{Physical Model}
\label{sec.3.2.1}

\begin{figure}[t]

\begin{flushleft} (a)\\ \end{flushleft} 
\includegraphics[width=.85\linewidth]{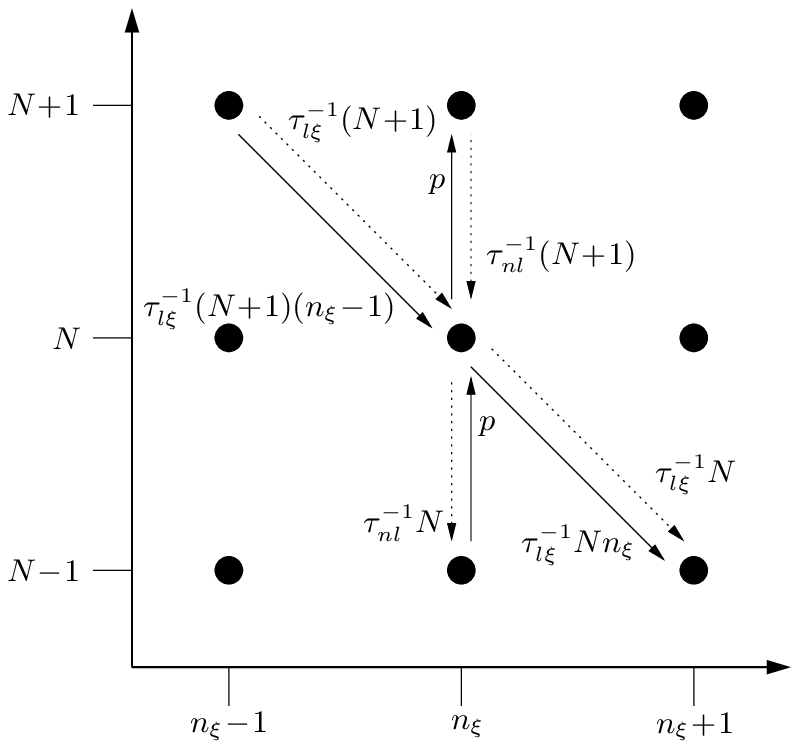} 
\begin{flushleft} (b)\\ \end{flushleft} 
\includegraphics[width=.85\linewidth]{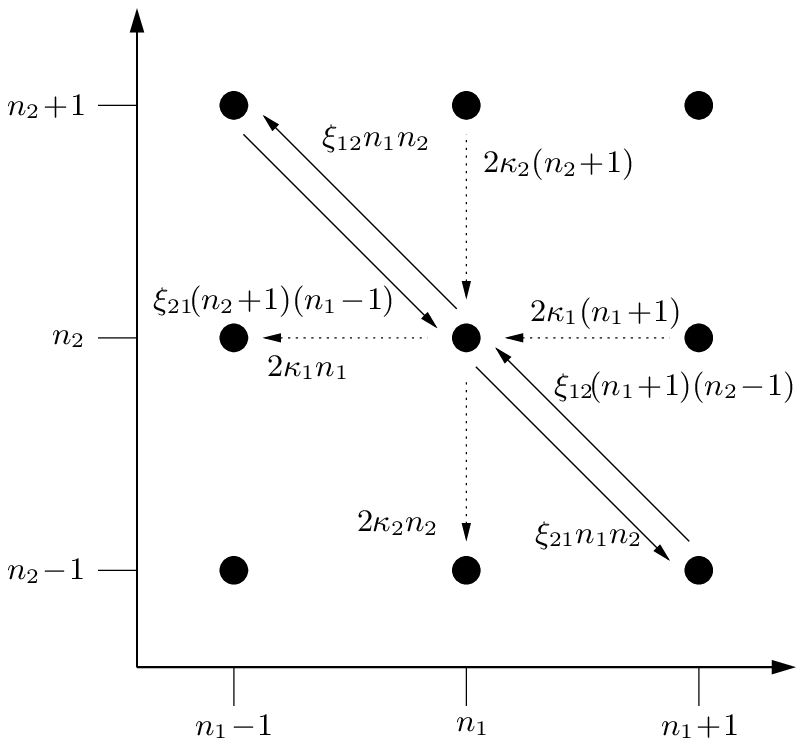}
\caption{\label{materiediagramm} Schematic representation of the
  various processes in the extended birth-death-model. The upper
  sketch (a) illustrates the light-matter-interaction and shows 
the transition rates into and out of the state with photon
number $n_\xi$ and carrier number $N$. Solid arrows represent stimulated
emission and pump, dotted arrows show spontaneous emission.
The lower sketch (b) illustrates
the interaction of the modes and shows 
the transition rates into and out of the state with photon
number $n_1$ and $n_2$. Solid arrows show mode interaction,
dotted arrows represent cavity losses.} 
\end{figure}

To arrive at the final form of the master equation a birth and death
model, analogue the one introduced by Rice and
Carmichael~\cite{Rice:4318}, is  considered. Transition rates into and
out of the state $\rho^{n_1,n_2}_N$ are connected to the relevant
processes in the coupled photon carrier
system. Figure~\ref{materiediagramm} shows how the master equation is
derived on a phenomenological level. The filled circles represent a state with
$N$ excited carriers, $n_1$ and $n_2$ photons in the cavity modes,
i.e. the diagonal elements $\rho^{n_1,n_2}_N$ of the density
matrix. The photon distribution for mode $\xi$ is
gained by summation over the remaining indices
$P(n_\xi)$ $\!=$ $\!\sum_{N,n_\zeta}
\rho^{n_\xi,n_\zeta}_N$. Figure~\ref{materiediagramm}(a) illustrates the
coupling of one mode to the gain medium. The horizontal axis shows the
number of photons $n_{\xi}$ in mode $\xi$ and the vertical axis
shows the number of excited carriers $N$. The carrier generation is
represented by solid vertical arrows since the photon number is not
changed. The rate of carrier generation in the excited level is given
by the pump power $p$. Vertical dotted arrows indicate the loss of excited
carriers due to spontaneous emission into non lasing modes. Moreover, the
emission into the cavity modes is represented by pairs of diagonal arrows
corresponding to spontaneous (dotted arrow) and stimulated (solid
arrow) emissions, since an excited carrier is lost and one photon in
one of the modes is gained. 
The factors $\tau _{l1}^{-1}$ and $\tau _{l2}^{-1}$ are introduced,
which represent the fractions of laser emission rate into the cavity
modes $1$ and $2$, correspondingly, where the relation
$\tau _{l1}^{-1} + \tau_{l2}^{-1} = \tau _{l}^{-1}$ holds.
Further, the interaction of the modes is
illustrated in Fig.~\ref{materiediagramm}(b). The two axes show the
number of photons $n_1$, $n_2$ in the modes $1$ and $2$,
respectively. The horizontal and vertical dotted arrows represent the
cavity losses of the two lasing modes with the loss rates $2\kappa_{\xi}$. 
%
A phenomenological nonlinear coupling between the lasing modes $1$ and
$2$ mediated by the gain medium is also introduced.  
In contrast to the Microscopic Semiconductor Theory (see
Sec.~\ref{sec.3.1}), where the coupling between the cavity modes is
mediated by the overlap of the mode functions with the gain carriers, here a 
nonlinear coupling between the lasing modes $1$ and $2$ is introduced
phenomenologically. 
The mode coupling strengths 
$\xi_{12}$ and $\xi_{21}$ are represented by the diagonal solid arrows
in the sketch and play the role of the detuning between the modes used in 
Sec.~\ref{sec.3.1}.
The complete master equation derived by the phenomenological birth and
death model reads: 
\begin{multline}
  \label{3.2.2}
  \frac{d}{dt}\rho^{n_1,n_2}_N=p\left[\rho^{n_1,n_2}_{N-1}-\rho^{n_1,n_2}_N\right]
    \\[1ex] -\tau _{nl}^{-1}[N\rho^{n_1,n_2}_N-(N+1)\rho^{n_1,n_2}_{N+1}]
  \\[1ex] -\tau _{l1}^{-1}[n_1N\rho^{n_1,n_2}_N-(n_1-1)(N+1)\rho^{n_1-1,n_2}_{N+1}]
    \\[1ex] -\tau _{l2}^{-1}[n_2N\rho^{n_1,n_2}_N-(n_2-1)(N+1)\rho^{n_1,n_2-1}_{N+1}]
     \\[1ex] -\tau _{l1}^{-1}[N\rho^{n_1,n_2}_N-(N+1)\rho^{n_1-1,n_2}_{N+1}]
    \\[1ex] -\tau _{l2}^{-1}[N\rho^{n_1,n_2}_N-(N+1)\rho^{n_1,n_2-1}_{N+1}]
      \\[1ex]-2\kappa_{1}[n_{1}\rho^{n_1,n_2}_N-(n_1+1)\rho^{n_1+1,n_2}_N]
  \\[1ex] -2\kappa_{2}[n_{2}\rho^{n_1,n_2}_N-(n_2+1)\rho^{n_1,n_2+1}_N]
    \\[1ex] -\xi_{12}[n_1n_2\rho^{n_1,n_2}_N-(n_1+1)(n_2-1)\rho^{n_1+1,n_2-1}_{N}]
    \\[1ex] -\xi_{21}[n_1n_2\rho^{n_1,n_2}_N-(n_1-1)(n_2+1)\rho^{n_1-1,n_2+1}_{N}].
  \end{multline}
In the above, each line corresponds to a process in the coupled
carrier photon system, i.e. to arrows going in and out of a state
$\rho^{n_1,n_2}_N$ in Fig.~\ref{materiediagramm}.

\subsubsection{Results}
\label{sec.3.2.2}
\begin{figure}[t]
\begin{flushleft} (a)\\ \end{flushleft}
\includegraphics[width=.9\linewidth]{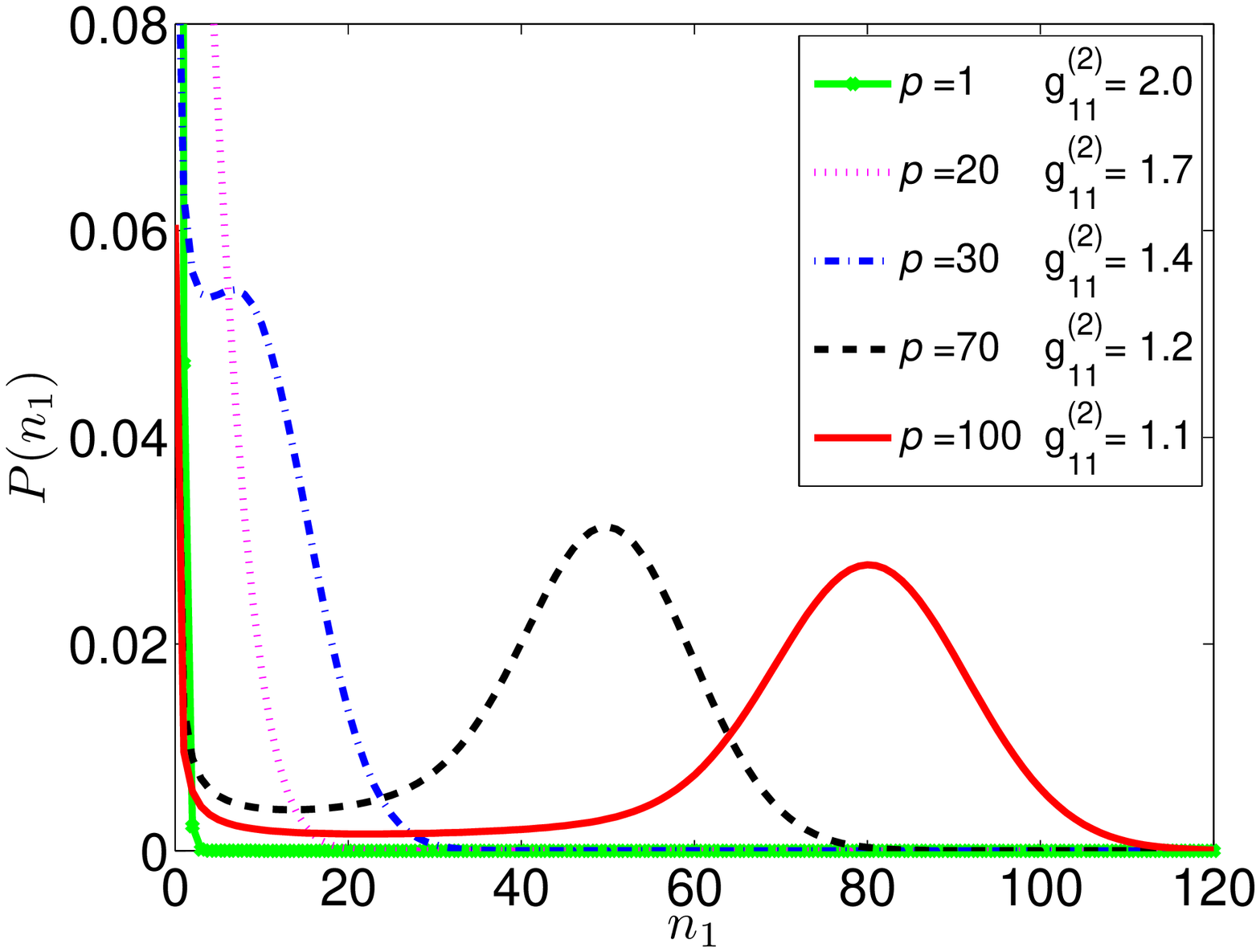}
\\
\begin{flushleft} (b)\\ \end{flushleft}
\includegraphics[width=.9\linewidth]{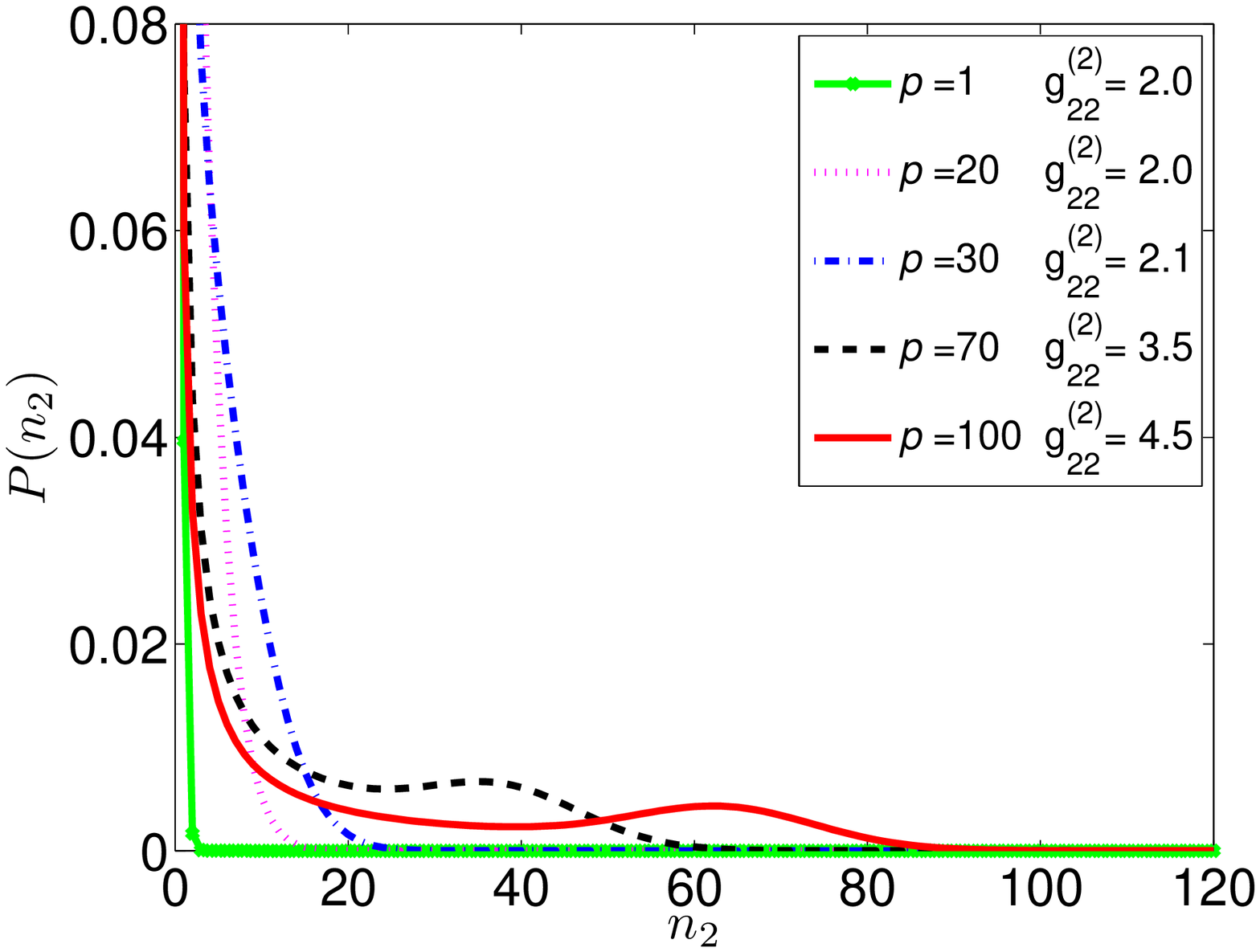}
\caption{\label{photondistrib} (color online) Photon probability
  distributions for mode 1 (a) and mode 2 (b) 
are shown 
for
 different pump rates in the units of $[\tau _{sp}^{-1}]$ and for $\tau _{l1}^{-1}$ $\!=$ $\!\tau
 _{l2}^{-1}$ $\!=$ $\!0.05$ $\!\tau _{sp}^{-1}$, $\tau _{nl}^{-1}$ $\!=$
 $\!0.9$ $\!\tau _{sp}^{-1}$, $2\kappa_{1}$ $\!=$ $\!\tau 
 _{sp}^{-1}$, $2\kappa_{2}$ $\!=$ $\!1.2$ $\!\tau _{sp}^{-1}$, $\xi_{12}$ $\!=$
 $\!\xi_{21}$ $\!=$ $\!0.1$ $\!\tau _{sp}^{-1}$.  
}
\end{figure}

The stationary solution of Eq.~(\ref{3.2.2}) gives the photon
probability distribution. Figure~\ref{photondistrib} shows the photon
distributions $P(n_\xi)$ for various pump strengths. Above the laser
threshold the autocorrelation functions of the modes $1$ and $2$ are
quite different, while $g^{(2)}_{11}(0)$ drops to values close to one
indicating Poissonian statistics, $g^{(2)}_{22}(0)$ rises up to values
substantially larger then two (thermal statistics). The results for
the autocorrelation functions of the modes $1$ and $2$ are in full
agreement with the ones obtained within the Microscopic Semiconductor
Theory in~Sec.~\ref{sec.3.1} (see e.g. Fig.~\ref{fig3}). However, in
contrast to the experimental data presented in Fig.~\ref{fig2}, the
autocorrelation function of the mode $2$ monotonically increases also
for high pump power rates.

The full photon
statistics reveals that both mode statistics exhibit a double peak
structure. The first peak appears at the zero photon state declining 
very steep and a second Poissonian like peak appears at the higher
photon states. In the mode $1$ the Poissonian peak is very pronounced
and dominates the statistics, the mode $2$ is dominated by the first
peak at zero photon number states. These statistics combined with the
fact of a crosscorrelation function far below unity allows for the
interpretation of a switching behavior of the modes. Both modes are in
a superposition of a lasing and a non lasing state and are alternating
in between them. 

\section{Discussion and Conclusions}
\label{sec.4}

We have investigated 
laser emission of electrically pumped quantum dots in a bimodal
micropillar cavity with special emphasis to the effects induced by
gain competition of the two orthogonally polarized modes.

The system consisting of a single low-density layer of QDs
and two spectrally splitted but overlapping modes with nearly equal
$Q$-factors, induced from the double degenerate fundamental mode by
slight cross-section asymmetry of the pillar, represents a viable
platform for the study of the coupling of two cavity modes in the
presence of a common gain medium. 
The polarization resolved measurements of the statistical properties
of the emitted light reveal, that the two competing modes display
completely different features. 
One of the modes (mode $1$) demonstrates typical statistical behavior
of a laser mode, namely the mode intensity displays the usual
"s"-shaped input-output characteristic, and  
the autocorrelation function
at zero time delay, measured using a 
HBT setup,
indicates the transition from spontaneous to
stimulated emission for increasing pump rates.
The measurements of the input-output characteristic of mode $2$ indicate the
threshold behavior, but for further increasing pump
rates the intensity saturates and even decreases, as the result of the
competition of the two modes induced by the common gain material.
Moreover, the autocorrelation function at zero time delay of mode $2$
at certain pump rates higher than the threshold values exhibits 
intensity fluctuation much higher than for a thermal state.
It is worth to note, that at these rates of the injection current the
anti-correlation between the two modes is the strongest.
For even larger pump rates, the crosstalk between the modes induces a
reduction of the autocorrelation function at zero time delay of the
mode $2$ reaching the value for a lasing mode.
Similarly, at these pump rates the crosscorrelation measurements
indicate increasing correlation between the modes due to the
crosstalk.   
The experimental results have been supported by the theoretical
calculations within the framework of the Microscopic Semiconductor
Theory~\cite{Gies:013803}, which we have extended to the case of two cavity modes
interacting with the QD-gain medium. Using the
cluster expansion scheme for correlation functions, we have obtained
the emission statistics of the carrier-photon system in the bimodal
cavity, taking into account the many-body effects. Importantly, within
our approach the effects related to the coupling
of the two modes of the bimodal cavity, induced by the interaction
with the common QD carriers are consistently included on the
microscopic level. The solution of the equations of motion for
correlation functions reveals, that indeed the autocorrelation function
of the mode $2$ for the pump rates larger than the threshold rate
reaches values well above $g^{(2)}(0)=2$, that corresponds to the thermal
state of light. The decrease of the crosscorrelation function of the
two modes below unity indicates anti-correlated
behavior of the mode coupling at these pump rates.
In fact, this effect can be explained by random intensity switching between the
two modes, which has negligible
influence on the photon statistics of the lasing mode,
but strongly affects the mode $2$ for which the relative strength of
fluctuations is larger.
It is worth to mention, that 
in the case of macroscopic two-mode ring lasers~\cite{singh:2459}
large intensity fluctuations have been
also found in the statistics of the more lossy mode, as the result of
the mode competition with the favored mode and emission switching of
the common atomic ensemble.     

To complement the theoretical results with the photon-number
statistics and to provide an
interpretation to the super-thermal intensity fluctuations
of mode $2$, we
have extended the birth-death model of the master
equation to the case of a bimodal cavity. In particular, we have
assumed a phenomenological nonlinear coupling between the cavity
modes, mediated by the overlap of the modes with the gain carriers. In
this way, solving the complete master equation, we have shown, that
the photon number statistics of the both modes exhibit similar double
peak structure---a peak at the zero photon state and a second peak at
a higher photon number. The results imply, that the both modes are in
a superposition of a lasing state and a thermal-like state. Whereas, the Poisson
peak at the higher photon number dominates the statistics of the mode
$1$, the statistics of the mode $2$ indicates 
thermal
state-like behavior, with a pronounced peak at zero photon state
complemented with a local maximum at a higher photon number
state. Thus, we may conclude that the photon number distribution of
the two modes approves the switching behavior of the interaction of
the two modes with the common gain medium. 

Similar double peak curve has been also observed 
in a semiclassical approach
for the intensity
probability distributions of the both modes of ring lasers~\cite{singh:2459},
where the nonzero values of the light intensity are much more probable
than the zero values for the favored mode and the other way around for
the lossy mode.  Moreover, a double peak structure of the photon
number distribution has been found for the composite mode at threshold
in the two-mode open laser theory~\cite{eremeev:023816}, where both
modes interact with the common ensemble of atoms and with the common
dissipation system.


\section{Acknowledgment}

This work was financially supported by the Deutsche
Forschungsgemeinschaft within the research grants RE2974/2-1 and
Wi1986/3-1 and the State of Bavaria. The authors gratefully thank
M.~Emmerling and A.~Wolf for expert sample preparation. A.~Foerster
acknowledges financial support from the 
Graduiertenf\"orderung Sachsen-Anhalt.


\appendix

\section{Equations for Microscopic Semiconductor Model }

In this Appendix we present the equations of motion
that
together with Eqs.~(\ref{3.6}), (\ref{3.7}), (\ref{3.9}), (\ref{3.10}) and
(\ref{3.12}) complete the full set of equations of motion for
one-time correlation functions on the quadruplet level of the cluster
expansion:
\label{app:1}
\begin{multline}
  \label{app.7}
  \frac{d}{dt} \delta \!\langle c^\dagger_j
  v_jb_\xi\rangle=
  i
  (\Delta_{i\xi }+i \kappa_\xi-i\Gamma)
  \delta \! \langle c^\dagger_j v_j b_\xi \rangle
\\[.5ex]
  +g_{\xi j}  \delta \! \langle c^\dagger_j c_j \rangle
  (1- \delta \!\langle v^\dagger_j v_j \rangle )
+
 \sum_{\zeta}
  \left[
    g_{\zeta j}
    \delta \! \langle b^\dagger_\zeta b_\xi \rangle
    (
    \delta \! \langle c^\dagger_j c_j \rangle
    -\delta \!\langle v^\dagger_j v_j \rangle
    )
  \right.
 \\[.5ex]
  \left.
    +g_{\zeta j}  \delta \!\langle c^\dagger_j c_j b^\dagger_\zeta b_\xi \rangle
    -g_{\zeta j}  \delta \! \langle v^\dagger_j v_j b^\dagger_\zeta b_\xi \rangle
 \right] ,
\end{multline}

\begin{multline}
  \label{app.9}
  \frac{d}{dt} \delta \!\langle v^\dagger_s v_s\rangle
  =\left( \sum_\xi g_{\xi j}  \delta \! \langle c^\dagger_s v_s b_\xi \rangle
    +\mathrm{H.c.}\right )
\\[.5ex]
  -
  \delta \! \langle v^\dagger_p , v_p \rangle
  (1- \delta \! \langle v^\dagger_s , v_s \rangle)
  \tau_v^{-1}
\!+
  \delta \! \langle c^\dagger_s c_s \rangle
  (1- \delta \! \langle v^\dagger_sv_s \rangle)
  \tau_{nl} ^{-1},
\end{multline}

\begin{multline}
  \label{app.10}
  \frac{d}{dt} \delta \!\langle v^\dagger_p v_p\rangle
  =-P( \delta \! \langle v^\dagger_p  v_p\rangle- \delta \! \langle
  c^\dagger_p  c_p\rangle )
\\[.5ex]
  +
  \delta \! \langle v^\dagger_p , v_p \rangle
  (1- \delta \! \langle v^\dagger_s , v_s \rangle)
  \tau_v^{-1}
  \!+
  \delta \! \langle c^\dagger_p c_p \rangle
  (1- \delta \! \langle v^\dagger_pv_p \rangle)
  \tau_{sp} ^{-1}
   ,
\end{multline}

\begin{multline}
  \label{app.11}
  \frac{d}{dt} \delta \!\langle c^\dagger_j c_j b^\dagger_\xi b_\zeta \rangle
  =- (\kappa_\xi + \kappa_\zeta )\delta \! \langle c^\dagger_j c_j b^\dagger_\xi b_\zeta \rangle
  \\
  -g_{\xi j} \delta \! \langle c^\dagger_j c_j \rangle
  \delta \! \langle
  c^\dagger_j v_j b_\zeta \rangle
  -
  g_{\zeta j} \delta \! \langle c^\dagger_j c_j \rangle
  \delta \! \langle v^\dagger_j c_j b^\dagger_\xi \rangle
  \\
  -\sum_{\xi '} \left(g_{\xi 'j}  \delta \! \langle c^\dagger_j v_j
    b^\dagger_\xi b_{\xi '} b_\zeta \rangle
  -g_{\xi 'j}  \delta \! \langle c^\dagger_j v_j b_\zeta \rangle  \delta \!
  \langle b^\dagger_\xi b_{\xi '} \rangle
  \right.
  \\
  \left.
    -g_{\xi 'j}  \delta \! \langle v^\dagger_j c_j b^\dagger_{\xi '} b^\dagger_\xi b_\zeta
    \rangle
    -g_{\xi 'j}  \delta \! \langle v^\dagger_j c_j b^\dagger_\xi \rangle
    \delta \! \langle b^\dagger_{\xi '} b_\zeta \rangle
\right)
,
\end{multline}
\begin{multline}
  \label{app.12}
  \frac{d}{dt} \delta \!\langle v^\dagger_j v_j b^\dagger_\xi b_\zeta\rangle
  =-(\kappa_\xi + \kappa_\zeta )  \delta \! \langle v^\dagger_j v_j b^\dagger_\xi b_\zeta \rangle
  \\[.5ex]
  +\sum_{\xi '}
  \left[
  g_{\xi 'j}  \delta \! \langle c^\dagger_j v_j b^\dagger_\xi b_{\xi '} b_\zeta \rangle
  +g_{\xi 'j}  \delta \! \langle c^\dagger_j v_j b_\zeta \rangle
  (1 -  \delta \! \langle v^\dagger_j v_j \rangle + \delta \! \langle
  b^\dagger_\xi b_{\xi '} \rangle)
  \right.
  \\
\left.
  +g_{\xi 'j}  \delta \! \langle v^\dagger_j c_j b^\dagger_{\xi '} b^\dagger_\xi b_\zeta \rangle
  +g_{\xi 'j}  \delta \! \langle v^\dagger_j c_j b^\dagger_\xi \rangle(1
  -\delta \! \langle v^\dagger_j v_j \rangle + \delta \! \langle
  b^\dagger_{\xi '} b_\zeta \rangle)
  \right]
  ,
\end{multline}
\begin{multline}
  \label{app.13}
  \frac{d}{dt} \delta \!\langle c^\dagger_j v_j b^\dagger_\xi b_\zeta
  b_{\xi '}\rangle
\\
  =i[\Delta _{i\xi }  +i (\kappa_\xi + \kappa_\zeta +\kappa_{\xi '})  +i\Gamma  ]  \delta \! \langle
  c^\dagger_j v_j b^\dagger_\xi b_\zeta b_{\xi '} \rangle
  \\
  -g_{\xi ' j}  \delta \! \langle c^\dagger_j c_j \rangle ( \delta \! \langle
  v^\dagger_j v_j b^\dagger_\xi b_\zeta \rangle
  - \delta \! \langle v^\dagger_j v_j b^\dagger_\xi b_{\xi '} \rangle
  + \delta \! \langle b^\dagger_{\zeta '} b^\dagger_\xi b_\zeta b_{\xi '} \rangle )
\\
+
\sum _{\zeta '}
\left[
g_{\zeta ' j}  \delta \! \langle c^\dagger_j c_j b^\dagger_\xi b_\zeta \rangle(1-
\delta \! \langle v^\dagger_j v_j \rangle + \delta \! \langle
b^\dagger_{\zeta '} b_{\xi '} \rangle)
\right.
\\
+g_{\zeta ' j}  \delta \! \langle c^\dagger_j c_j b^\dagger_\xi b_{\xi '} \rangle(1 -
\delta \! \langle v^\dagger_j v_j \rangle - \delta \! \langle
b^\dagger_{\zeta '} b_\zeta \rangle)
\\
-2 g_{\zeta ' j}  \delta \! \langle c^\dagger_j v_j b_\zeta \rangle  \delta \!
\langle c^\dagger_j v_j b_{\xi '} \rangle
-g_{\zeta ' j}  \delta \! \langle v^\dagger_j v_j \rangle  \delta \! \langle
b^\dagger_{\zeta ' j} b^\dagger_\xi b_\zeta b_{\xi '} \rangle
\\
\left.
-g_{\zeta ' j}  \delta \! \langle v^\dagger_j v_j b^\dagger_\xi b_\zeta \rangle
\delta \! \langle b^\dagger_{\zeta '} b_{\xi '} \rangle
-g_{\zeta ' j}  \delta \! \langle v^\dagger_j v_j b^\dagger_\xi b_{\xi '} \rangle
\delta \! \langle b^\dagger_{\zeta '} b_\zeta \rangle
\right]
,
\end{multline}
\begin{multline}
  \label{app.14}
  \frac{d}{dt} \delta \!\langle v^\dagger_j c_j b^\dagger_\xi
  b^\dagger_\zeta b_{\xi '}\rangle
\\
  =i[-\Delta _{iq} + i (\kappa_\xi + \kappa_\zeta +\kappa_{\xi '}) +i\Gamma  ]  \delta \! \langle
  v^\dagger_j c_j b^\dagger_\xi b^\dagger_\zeta b_{\xi '} \rangle
\\
 -g_j  \delta \! \langle c^\dagger_j c_j \rangle  (\delta \! \langle
 v^\dagger_j v_j b^\dagger_\xi b_{\xi '} \rangle -  \delta \! \langle
 v^\dagger_j v_j b^\dagger_\zeta b_{\xi '} \rangle + \delta \! \langle
 b^\dagger_\xi b^\dagger_\zeta b_n b_{\xi '} \rangle)
\\
\sum _{\zeta '}
\left[
+g_{\zeta ' j}  \delta \! \langle c^\dagger_j c_j b^\dagger_\xi b_{\xi '} \rangle(1 -
\delta \! \langle v^\dagger_j v_j \rangle +  \delta \! \langle
b^\dagger_\zeta b_{\zeta '} \rangle )
\right.
\\
+g_{\zeta ' j}  \delta \! \langle c^\dagger_j c_j b^\dagger_\zeta b_{\xi '} \rangle(1 -
\delta \! \langle v^\dagger_j v_j \rangle + \delta \! \langle
b^\dagger_\xi b_{\zeta '} \rangle)
\\
-2 g_{\zeta ' j}  \delta \! \langle v^\dagger_j c_j b^\dagger_\xi \rangle  \delta
\! \langle v^\dagger_j c_j b^\dagger_\zeta \rangle
-g_{\zeta ' j}  \delta \! \langle v^\dagger_j v_j \rangle  \delta \! \langle
b^\dagger_\xi b^\dagger_\zeta b_{\zeta '} b_{\xi '} \rangle
\\
\left.
-g_{\zeta ' j}  \delta \! \langle v^\dagger_j v_j b^\dagger_\xi b_{\xi '} \rangle
\delta \! \langle b^\dagger_\zeta b_{\zeta '} \rangle
-g_{\zeta ' j}  \delta \! \langle
v^\dagger_j v_j b^\dagger_\zeta b_{\xi '} \rangle  \delta \! \langle b^\dagger_\xi
b_{\zeta '} \rangle
\right]
.
\end{multline}

\bibliography{Exp}

\end{document}